\documentclass[aps,pre,groupedaddress,showpacs,preprint]{revtex4}
\usepackage{graphicx}% Include figure files
\usepackage[dvips]{epsfig}% Include figure files
\usepackage{dcolumn}% Align table columns on decimal point
\usepackage{subfigure}%
\usepackage{bm}% bold math
\usepackage{amssymb}
\usepackage{amsmath}

\begin{document}

\title{Stochastic growth of radial clusters: weak convergence to
the asymptotic profile and implications for morphogenesis}

\author{Carlos Escudero}

\affiliation{Departamento de Econom\'{\i}a Cuantitativa \& \\
Instituto de Ciencias Matem\'aticas (CSIC-UAM-UC3M-UCM), \\
Universidad Aut\'onoma de Madrid, \\
Ciudad Universitaria de Cantoblanco, 28049 Madrid, Spain}

\begin{abstract}
The asymptotic shape of randomly growing radial clusters is
studied. We pose the problem in terms of the dynamics of
stochastic partial differential equations. We concentrate on the
properties of the realizations of the stochastic growth process
and in particular on the interface fluctuations. Our goal is
unveiling under which conditions the developing radial cluster
asymptotically {\it weakly} converges to the concentrically
propagating spherically symmetric profile or either to a symmetry
breaking shape. We demonstrate that the long range correlations of
the surface fluctuations obey a self-affine scaling and that scale
invariance is achieved by means of the introduction of three
critical exponents. These are able to characterize the large scale
dynamics and to describe those regimes dominated by system size
evolution. The connection of these results with mathematical
morphogenetic problems is also outlined.
\end{abstract}

\pacs{05.40.-a, 05.65.+b, 87.10.-e, 87.10.Mn}

\maketitle

\section{Introduction}

The building blocks of mathematical morphogenesis were put several
decades ago in the seminal works of Thompson~\cite{thompson} and
Turing~\cite{turing}. A particularly relevant problem in this
context is the examination of the properties relating to the
architecture of cell colonies, as already noted and investigated
by Eden~\cite{eden1,eden2}. At some basic underlying level, one
could say that Turing and Eden shared the goal of understanding
how a macroscopic structure, in particular one breaking the
initial homogeneity, could arise out of a multiplicity of simple
interactions. In the introduction of~\cite{turing} one can read
Turing thoughts on how a structure composed by randomly and
isotropically proliferating cells could or could not break the
initial spherical symmetry. But then his approach moved to the use
of reaction--diffusion equations and to the search of a
deterministic mechanism, the currently well known Turing
instability, able to give rise to pattern formation. On the other
hand, Eden focused his interest on the evolution of a pure growth
process. He concentrated on a probabilistic abstraction of a
developing cell colony and studied stochastic symmetry breaking
concomitant to growth. In particular, Eden studied the
architecture of a lattice cell colony to which new cells were
added following certain probabilistic rules. The objective was
determining the asymptotic colony profile. If the growth rules are
isotropic then the realizations of the growth process are either
spherically symmetric or this spherical symmetry is broken by
means of long range fluctuations.

These sorts of discrete models are usually studied using continuum
equations for ease of analytical
treatment~\cite{vvedenky1,vvedensky2}. While most of these
developments have been carried out for systems in which there is
an external input of mass, the universality of the surface
fluctuations is expected to remain unchanged when the origin of
mass appearance is internal. In fact, the Eden model in a strip
geometry has been satisfactorily analyzed within the classical
theoretical framework~\cite{barabasi}. And, as we already have
mentioned, this model was introduced as a probabilistic
abstraction of a proliferating cell colony. So we expect our
results to hold independently of the internal or external origin
of mass input into the system. This will be so at least in those
cases in which the growing radial interface can be described as a
Monge patch in spherical coordinates. Growing clusters as the ones
generated by diffusion--limited aggregation processes yield more
complex structures that cannot be described in such simple terms,
and thus fall beyond the scope of the present approach.

Indeed, the original Eden problem can be greatly generalized by
means of the use of stochastic partial differential equations.
They allow a systematic study of the properties of the colony
periphery, particularly of the interface fluctuations. At the same
time, they allow us to get rid of the undesirable lattice
anisotropy. In this work we will concentrate on the properties of
the realizations of the stochastic growth process. Our goal is
unveiling under which conditions the developing radial cluster
asymptotically {\it weakly} converges to the concentrically
propagating spherically symmetric profile. Let us emphasize that
we are interested in determining the presence or absence of
spherical symmetry in every realization of the stochastic growth
process in the long time limit. As we are considering isotropic
growth which is free from deterministic instabilities, averaging
the cluster profile over many realizations yields an immediate
spherical symmetry. The properties of the realizations can be
ascertained, as we will see, calculating suitable correlations
adapted from classical elements of stochastic growth
theory~\cite{barabasi}, and keeping in mind that we should expect
weak rather than strong convergence of the asymptotic profiles.
Weak convergence is a mathematically totally precise concept which
physically corresponds to the convergence resulting from the
self-averaging of wildly oscillating quantities or
structures~\cite{foot}.

Apart from the overall shape, the microscopic fluctuations of
growing radial clusters have been studied as well. It turns out
that these fluctuations sculpture a fractal surface which is
statistically self-affine~\cite{barabasi,halpin,hammersley}. This
microscopic roughness, in those cases in which the pointwise
width~\cite{width} is well defined (which actually rules out all
flat interfaces due to the microscopic properties of white noise)
is described by the familiar scaling of planar
systems~\cite{family}. This is not so, however, for long range
fluctuations~\cite{escudero}. Long range radial correlations
cannot be in general deduced from planar scalings. They still
adopt a self-affine form for macroscopic length scales, but scale
invariance is this time achieved for different values of the
critical exponents. The following sections will discuss the
numerical values and physical meaning of these exponents. We will
see that they are able to characterize which regimes
asymptotically weakly converge to the radially symmetric shape and
which do not.

\section{Radial correlations}

\begin{figure}[h]
\begin{center}
\includegraphics[scale=0.5]{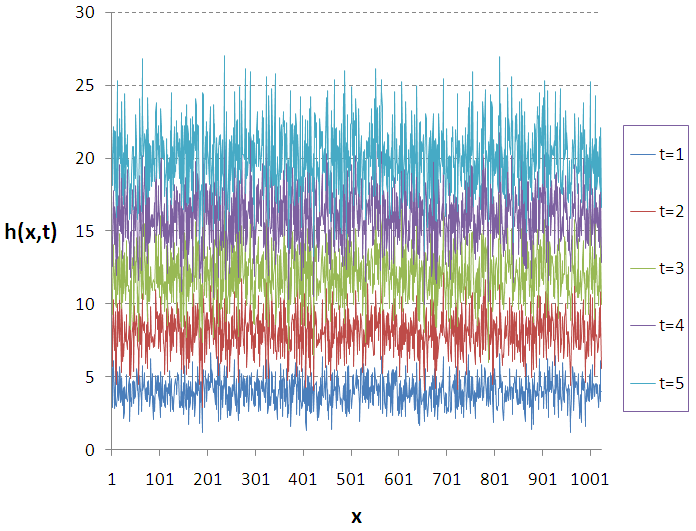}
\caption{Solution to the random deposition
equation~(\ref{rdequation}) at several times. The values of the
parameters are $F=4$, $\epsilon=1$, and the system has been
discretized in $2^{10}$ spatial points. We have used as initial
condition $h(x,0)=0$.} \label{rdfig}
\end{center}
\end{figure}

In planar situations of non-equilibrium growth the interface is
parameterized by the height function $h(x,t)$, which indicates the
displacement of the interface with respect to some hyperplane
taken as origin, at some $d-$dimensional spatial point $x$ and for
some instant of time $t$~\cite{barabasi}. We illustrate this
geometry in Fig.~(\ref{rdfig}), where we present the
one-dimensional solution to the so-called random deposition
equation
\begin{equation}
\label{rdequation}
\partial_t h= F + \xi(x,t),
\end{equation}
where $\xi$ is a zero-mean Gaussian noise whose correlation is
$\left< \xi(x,t)\xi(x',t) \right>= \epsilon
\delta(x-x')\delta(t-t')$. This numerical approximation is found
by means of spatially discretizing the system.

\begin{figure}[h]
\begin{center}
\includegraphics[scale=0.7]{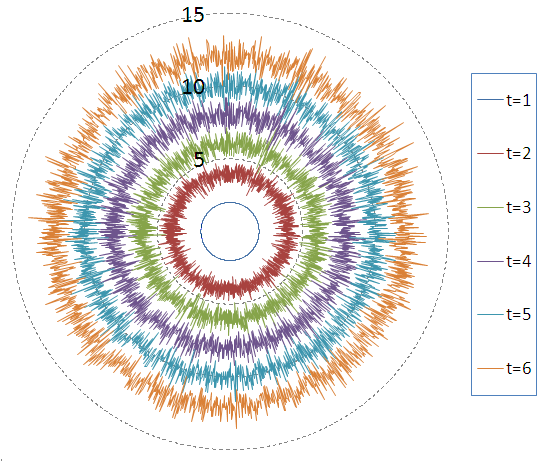}
\caption{Radial random deposition growth process plotted at
different times. The values of the parameters are $F=2$,
$\gamma=1$, $\epsilon=(200\pi)^{-1}$, and the system has been
discretized in $2^{11}$ spatial points. We have used as initial
condition $r(\theta,1)=2$ (note that we have used as initial time
$t_0=1$).} \label{rrdfig}
\end{center}
\end{figure}

In these systems and for large spatiotemporal scales the
height-height correlation function adopts the self-affine scaling
form
\begin{eqnarray}
\nonumber \left< h(x,t)h(x',t) \right> &=& t^{2 \beta_c} f \left(
\frac{|x-x'|}{t^{1/z_c}} \right)= \\
\label{fv} &=& |x-x'|^{2 \alpha_c} g \left(
\frac{|x-x'|}{t^{1/z_c}} \right),
\end{eqnarray}
which is invariant to the transformation $x \to b x$, $t \to b^z
t$ and $h \to b^\alpha h$, for $\alpha=\beta z$ and
$\alpha=\alpha_c$, $\beta=\beta_c$ and $z=z_c$, where the
subscript $c$ denotes the classical value of an exponent for a
given model. $\alpha$ is known as roughness exponent, $\beta$ as
growth exponent and $z$ as dynamic exponent, $f$ and $g$ are
scaling functions. This scaling is valid for a large class of
linear and nonlinear models. If we consider the linear Langevin
equation
\begin{equation}
\label{langevin}
\partial_t h= -D |\nabla|^\zeta h + \xi(x,t),
\end{equation}
where $\xi$, as before, is a zero-mean Gaussian noise whose
correlation is $\left< \xi(x,t)\xi(x',t) \right>= \epsilon
\delta(x-x')\delta(t-t')$ and $|\nabla|^\zeta$ is a
pseudo-differential operator to be understood in the Fourier
transform sense, the exponents read $z = \zeta$, $\alpha = (\zeta
-d)/2$ and $\beta = 1/2-d/(2 \zeta)$. One would like to know if
the radial counterpart of Eq.~(\ref{langevin}) displays an
analogous behavior to that dictated by scaling~(\ref{fv}). In the
radial extension of the theory one has a new degree of freedom
which was trivial in the planar case: the rate of growth. If we
assume that the average radius (measured as distance from the
origin) of the macroscopic radial form grows as a power law $
\left< r \right> = F t^\gamma$ for $t \ge t_0$ and a growth index
$\gamma
>0$, then the equation for the radial surface fluctuation
$\rho(\theta,t)$ is
\begin{equation}
\label{radeq}
\partial_t \rho = -\frac{D}{F^\zeta t^{\zeta \gamma}}|\nabla_\theta|^\zeta \rho + \frac{1}{F^{d/2} t^{\gamma
d/2}}\frac{\eta(\theta,t)}{J(\theta)^{1/2}},
\end{equation}
according to the reparametrization invariance
principle~\cite{marsili,escudero2}, and where $\theta$ denotes the
set of angles parameterizing the $d-$dimensional radial interface
and $J(\theta)$ is the Jacobian determinant of the change of
variables from $(x,h)$ to $(\theta,r)$ evaluated at
$r=1$~\cite{escudero4}. The noise is simply
$\eta=\xi/\sqrt{\epsilon}$. In order to visually compare planar
and radial processes we have plotted the field $r(\theta,t) =
\left\langle r \right\rangle + \sqrt{\epsilon} \rho(\theta,t)$ for
$D=0$ in Fig.~(\ref{rrdfig}). Neglecting the diffusion constant
makes this stochastic process the radial counterpart of the random
deposition process plotted in Fig.~(\ref{rdfig}). We will
therefore refer to it as the {\it radial random deposition
process}.

As we have mentioned in the introduction, one of the motivations
of this work is studying the large scale, long time properties of
a randomly proliferating set of cells or entities in general. Such
a process can be modelled in some detail using a master equation,
which can be mapped into a stochastic partial differential
equation provided with a multiplicative noise
term~\cite{emilio,carlos}. On the other hand, we focus in this
work on stochastic partial differential equations provided with
additive noise. While this could seem as an apparent paradox, the
underlying reason is that equations~(\ref{langevin})
and~(\ref{radeq}) are effective models. So they are intended to
describe just the long range properties of the growing interface,
and in this limit fluctuations can be effectively considered
additive. A nice example of this, a case in which a front driven
by multiplicative noise is found to belong to a universality class
described by an additive noise stochastic equation, is described
in~\cite{moro}.

The analysis of the exact solutions to Eq.~(\ref{radeq}) for
$\gamma>1/\zeta$ and long times shows the diffusion term is
irrelevant in the large scale $|\theta-\theta'| \gg
t^{1/\zeta-\gamma}$ and the correlation
becomes~\cite{escudero2,escudero4}
\begin{equation}
C(\theta,\theta',t) \sim \left\{
\begin{array}{lll} t^{1-\gamma d} \delta(\theta-\theta')/J(\theta) &
\mbox{\quad if \quad $\gamma d < 1$},
\\
\ln(t) \delta(\theta-\theta')/J(\theta) & \mbox{\quad if \quad
$\gamma d = 1$},
\\
\delta(\theta-\theta')/J(\theta) & \mbox{\quad if \quad $\gamma d
> 1$},
\end{array} \right.
\end{equation}
where $C(\theta,\theta',t)=\left< \rho(\theta,t) \rho(\theta',t)
\right>$. In terms of arc-length differences $\ell - \ell' \sim
t^\gamma (\theta-\theta')$ one finds~\cite{escudero4}
\begin{equation}
\label{radcorr1}
C(\ell,\ell',t) \sim \left\{
\begin{array}{lll} t \delta(\ell-\ell')/J(t^{-\gamma}\ell) &
\mbox{\quad if \quad $\gamma d < 1$},
\\
t \ln(t) \delta(\ell-\ell')/J(t^{-\gamma}\ell) & \mbox{\quad if
\quad $\gamma d = 1$},
\\
t^{\gamma d}\delta(\ell-\ell')/J(t^{-\gamma}\ell) & \mbox{\quad if
\quad $\gamma d
> 1$}.
\end{array} \right.
\end{equation}
This correlation is invariant to the transformation $\ell \to b
\ell$, $t \to b^{z} t$ and $\rho \to b^{\alpha} \rho$ for
$z=1/\gamma=:z_r$ and $\alpha=1/(2\gamma)-d/2=:\alpha_r$
($\alpha=0=:\alpha_r$) when $\gamma d<1$ ($\gamma d \ge 1$). We
define $\beta_r := \alpha/z= \alpha_r/z_r=(1-\gamma d)/2$
($\beta_r=0$) when $\gamma d<1$ ($\gamma d \ge 1$). First, the
functional form of this correlation is not reducible to scaling
form~(\ref{fv}), and second, the exponents are totally different
(this makes precise the analysis carried out in~\cite{escudero}).
In summary, long range radial surface fluctuations are not
describable in terms of scaling~(\ref{fv}).

\section{Dilution and universality class bifurcation}
\label{bifurcation}

It is clarifying to consider a simplified problem: an abstraction
in which Eq.~(\ref{langevin}) is placed on a growing
domain~\cite{escudero3}. If this is realized by means of the
transformation $x \to (t/t_0)^\gamma x$ one finds
\begin{equation}
\label{gdndil}
\partial_t h=-D\left( \frac{t_0}{t} \right)^{\zeta \gamma}
|\nabla|^\zeta h +\gamma F t^{\gamma-1}+\left( \frac{t_0}{t}
\right)^{d \gamma/2} \xi,
\end{equation}
where the mass source is explicitly included and we assume fast
growth $\gamma>1/\zeta$ and the rough interface inequality
$\zeta>d$. In this case it can be shown that the correlation obeys
scaling~(\ref{fv}) for short length scales $|x-x'| \ll t^{(1-\zeta
\gamma)/\zeta}$ and for long length scales $|x-x'| \gg t^{(1-\zeta
\gamma)/\zeta}$ it is~\cite{escudero3}
\begin{equation}
C(y,y',t) \sim \left\{
\begin{array}{lll} t \delta(y-y') &
\mbox{\quad if \quad $\gamma d < 1$},
\\
t \ln(t) \delta(y-y') & \mbox{\quad if \quad $\gamma d = 1$},
\\
t^{\gamma d}\delta(y-y') & \mbox{\quad if \quad $\gamma d
> 1$},
\end{array} \right.
\end{equation}
where $y=(t/t_0)^\gamma x$ is the counterpart of the arc-length
coordinate. This correlation is scale invariant for the exponents
$\{\alpha_r,z_r\}$ but not for the exponents $\{\alpha_c,z_c\}$.

This way of realizing domain growth agrees with reparametrization
invariance~\cite{escudero4}, but is not unique. It makes space and
mass grow simultaneously~\cite{escudero4}, while it is possible to
realize domain growth at constant mass by introducing a dilution
term~\cite{escudero3,crampin}
\begin{equation}
\partial_t h=-D\left( \frac{t_0}{t} \right)^{\zeta \gamma}
|\nabla|^\zeta h -\frac{d \gamma}{t}h +\gamma F
t^{\gamma-1}+\left( \frac{t_0}{t} \right)^{d \gamma/2} \xi.
\end{equation}
In this case the correlation is
\begin{equation}
\label{rd}
C(y,y',t) \sim t \delta(y-y'),
\end{equation}
which is scale invariant for both $\{\alpha_c,z_c\}$ and
$\{\tilde{\alpha}_r=1/(2\gamma)-d/2,z_r\}$. The first invariance
can be used to argue that this correlation is a direct consequence
of scaling~(\ref{fv}), but the scaling form~(\ref{rd}) is
definitely ambiguous in view of the second invariance. Both sort
of dynamics, with and without dilution, are physically realizable;
while dilution dynamics is apparently better suited to describe
some natural processes, no-dilution dynamics seems to agree better
with the numerical simulation of some discrete models as discussed
in~\cite{escudero4}. Together with the difference in the roughness
exponent, very pronounced differences in the auto-correlation and
persistence exponents separate the interface behavior in presence
and absence of dilution~\cite{escudero3}. This reveals that the
non-uniqueness of the extension of a Langevin equation to a
growing domain implies in turn a universality class bifurcation.

\section{Geometric constraints}

One can apply what has been learned in the previous section about
growing domains to the radial situation. Including dilution in the
radial equation for interface motion yields a different dynamics
from the one established by reparametrization invariance. The
correlation in this case becomes~\cite{escudero4}
\begin{equation}
\label{radcorr}
C(\ell,\ell',t) \sim t
\frac{\delta(\ell-\ell')}{J(t^{-\gamma}\ell)},
\end{equation}
which is scale invariant for the exponents
$\{\tilde{\alpha}_r,z_r\}$ but not for $\{\alpha_c,z_c\}$. The
difference among $\alpha_r$ and $\tilde{\alpha}_r$ originates in
the memory effects that affect the surface described by
reparametrization invariance and disappear by virtue of
dilution~\cite{escudero4,escudero3,escuderoadd}. The difference
between the sets $\{\alpha_r,z_r\}$ and
$\{\tilde{\alpha}_r,z_r\}$, and the classical set
$\{\alpha_c,z_c\}$ is of a different nature. Note that the
Langevin dynamics on a growing domain and influenced by dilution
gave rise to the random deposition correlation~(\ref{rd}):
\begin{equation}
\label{rd2}
\left\langle h(y,t) h(y',t) \right\rangle \sim t
\delta(y-y').
\end{equation}
This correlation is invariant to the transformation
\begin{equation}
\label{scale1}
y \to b y, \qquad t \to b^z t, \qquad \mathrm{and}
\qquad h \to b^\alpha h,
\end{equation}
for
\begin{equation}
\label{scale2}
z=\kappa \qquad \mathrm{and} \qquad \alpha=(\kappa - d)/2, \qquad
\kappa \in \mathbb{R};
\end{equation}
let us remark that scale invariance is achieved for any real
number $\kappa$, thus~(\ref{scale1}) and~(\ref{scale2}) actually
constitute a one-parameter family of scale transformations. This
is, there is one degree of freedom in the critical exponents, or
in other words, the scaling form is insensitive to the dynamic
exponent. Contrarily, in a radial geometry the correlation,
Eq.~(\ref{radcorr}), is scale invariant only if $\kappa=1/\gamma$:
this is the constraint imposed by the geometrical factor $J$. The
Jacobian determinant $J=\sin(\theta_1)^{d-1}\sin(\theta_2)^{d-2}
\cdots \sin(\theta_{d-1})$ for a $d-$dimensional interface, where
$\theta_1,\cdots,\theta_{d-1}$ is the set of polar angles, while
it is of course independent of the azimuth angle $\theta_d$. So
this constraint appears for $d \ge 2$, but it is not present for
$d=1$. The reason is that the one-dimensional radial interface is
topologically identical to the one-dimensional, topologically
toroidal, interface which results from assuming periodic boundary
conditions on the growing domain setting. In this case, the
angle-like coordinate $x$, to which the azimuth angle is akin, may
be either dimensional or dimensionless (this could be expressed by
means of the dichotomy of Lagrangian/Eulerian coordinates
similarly to the developments in~\cite{escudero3}). On the other
hand, the polar angles, ranging only $\pi$ radians, cannot be
associated to periodic coordinates which could have dimensions.
This fixes the scaling $z=1/\gamma$.

One can use the ambiguity of Eq.~(\ref{rd2}) to claim that
$z=\zeta$ is a good dynamic exponent in this case, but that would
be accepting that topology can modify the scaling. Also, the value
of the auto-correlation exponent in this case suggests that
$z=1/\gamma$ is the good dynamic exponent~\cite{escudero3}. In any
case, it is now clear that there is no basis for claiming that
correlation~(\ref{rd})-(\ref{rd2}) is a consequence of
scaling~(\ref{fv}) in the present framework. This is because,
although both forms are invariant to the
transformation~(\ref{scale1}) for $z=\zeta$ (for linear equations
like~(\ref{langevin})), self-similar form~(\ref{rd})-(\ref{rd2})
preserves scale invariance independently of the value of $z$ (as
specified in Eq.~(\ref{scale2})). So this second form is much more
general and the value $z=\zeta$ plays no special role in it (let
us note again that the value $z=1/\gamma$ plays a special role in
the sense that it appears in the auto-correlation
function~\cite{escudero3}). Despite this ambiguity in the growing
domain situation, it is clear that the radial
correlation~(\ref{radcorr}) is just invariant for
$\{\tilde{\alpha}_r,z_r\}$, and the classical exponents
$\{\alpha_c,z_c\}$ play no role in this case. This can be seen by
means of a direct application of transformation~(\ref{scale1}) to
this correlation and substituting $\alpha$ and $z$ for the two
different alternatives.

\section{Dimensional analysis and nonlinear dynamics}

Our results so far allow us to develop a consistent dimensional
analysis of radial equations. We start with Eq.~(\ref{radeq})
performing the scale transformation $\rho \to b^\alpha \rho$ and
$t \to b^z t$. The angular variables are dimensionless and this,
as specified by the factor $J$, fixes the exponent $z=1/\gamma$.
The stochastic term becomes scale invariant for
$\alpha=1/(2\gamma)-d/2$ and the diffusion irrelevant for $\gamma
> 1/\zeta$: this agrees with our previous analysis using the
explicit solutions of the linear equations. We can now use this
technique to get some insight on nonlinear equations. The
Kardar-Parisi-Zhang (KPZ) equation~\cite{kpz} for the radial
surface fluctuation can be written as
\begin{equation}
\partial_t \rho = \frac{\nu}{F^2 t^{2 \gamma}} \partial_\theta^2 \rho +
\frac{\lambda}{F^2 t^{2 \gamma}} (\partial_\theta \rho)^2 +
\frac{1}{F^{d/2} t^{\gamma
d/2}}\frac{\eta(\theta,t)}{J(\theta)^{1/2}}.
\end{equation}
We note that scaling arguments were successfully employed in
determining the critical behavior of KPZ with a time dependent
coefficient of the nonlinearity~\cite{ehg}. If we select the same
values of $\alpha$ and $z$ as in the linear case in order to make
the stochastic term scale invariant we find that diffusion and
nonlinearity are irrelevant for $\gamma > 1/2$ and $\gamma
> 3/(4+d)$ respectively. This agrees with the na\"{\i}f dimensional analysis of
KPZ and thus we cannot expect it to be correct. In fact, the
na\"{\i}f dynamical exponent $z=(4+d)/3$ found by means of a
dimensional analysis of the KPZ nonlinearity does not agree with
the one measured in simulations and successfully predicted by the
dynamic renormalization group analysis in one
dimension~\cite{barabasi}. It is plausible that the nonlinearity
interacts with the noise to yield a correlation different
from~(\ref{radcorr1}) (or from~(\ref{radcorr}) if dilution is
included). In order to guess what form the correlation would have
we can try to get some illumination from scaling~(\ref{fv}).
Taking the short time limit in~(\ref{fv}) we find
\begin{equation}
C(x,x',t) \sim t^{2 \beta_c +d/z_c} \delta(x-x'),
\end{equation}
where the temporal prefactor expresses the time dependence of the
variance of the interface center of mass
position~\cite{escudero4}. Using this form we modify, for
instance, Eq.~(\ref{radcorr}) to the new correlation
\begin{equation}
\label{radcorrkpz}
C(\ell,\ell',t) \sim t^{2 \beta_c +d/z_c}
\frac{\delta(\ell-\ell')}{J(t^{-\gamma}\ell)},
\end{equation}
which yields the exponents $\alpha=(2\beta_c
+d/z_c)/(2\gamma)-d/2$, $z=1/\gamma$ and $\beta_r=(2\beta_c
+d/z_c)/2-d\gamma/2$. In this case we have the diffusion is
irrelevant whenever $\gamma > 1/2$ and the nonlinearity is
irrelevant if $\gamma > 1/z_c$, where we have employed the
relation $\alpha_c + z_c = 2$ in the derivation of the last
inequality. This last inequality could be considered as desirable
but correlation~(\ref{radcorrkpz}) has a different drawback. As we
have said, the exponent $\beta_r$ describes the time power law
dependence of the standard deviation of the interface center of
mass position. For the Eden model, which in planar format belongs
to the KPZ universality class~\cite{barabasi}, $d=\gamma=1$ and so
$\beta_r = 1/6$. This exponent however does not agree with the one
measured in simulations~\cite{ferreira} and this constitutes
another disadvantage of the use of scaling~(\ref{fv}) in radial
systems. We note that considering the planar KPZ equation on a
growing domain does not simplify much things with respect to
considering the full radial case~\cite{wio}.

Together with the KPZ equation, the Villain-Lai-Das Sarma (VLDS)
equation~\cite{villain,lai} is one of most important nonlinear
models for surface growth. If one derives the VLDS equation in the
radial setting one finds irrelevance of the diffusion for $\gamma
>1/4$ and of the nonlinearity for $\gamma> 1/z_{c}$.
In this case the well known hyperscaling relation for conserved
growth models yields $2 \beta_{c} +d/z_{c}=1$, and so equality
of~(\ref{radcorrkpz}) with (\ref{radcorr}). This suggests that the
scaling properties derived herein for linear models are also valid
for some nonlinear models as the VLDS equation.

\section{Conclusions}

In this work we have examined the dynamic scaling of radial
interfaces. We have shown that long range fluctuations, which are
describable by two-point correlation functions, become scale
invariant for values of the critical exponents which may totally
differ from the corresponding ones in a planar geometry. In
rapidly growing radial systems we have found
\begin{equation}
\alpha =\frac{1}{2\gamma}-\frac{d}{2}, \quad z=\frac{1}{\gamma},
\quad \beta_r=\frac{1}{2}-\frac{\gamma d}{2},
\end{equation}
all of which have a clear physical meaning. The $\beta_r$
exponent, as we have discussed, describes the interface center of
mass fluctuations. The $\alpha$ exponent associated to the radial
surface fluctuation indicates how much the growing cluster
deviates from a $d-$dimensional growing hypersphere. For $\alpha <
0$ the growing cluster converges to a hypersphere concentrically
growing at the deterministic rate $t^\gamma$, and in this sense
the growth profile is flat. This sort of convergence is of course
{\it weak}: the profile converges to the hypersphere on (any type
of spatial) average. The microscopic roughness, which is given by
the height-difference correlation, reduces to that of the planar
profile~\cite{escudero3} and precludes strong or pointwise
convergence. This is so at least in situations in which the
height-difference correlation is well defined; otherwise a
different scaling is possible even locally in
space~\cite{escudero4}. This makes precise and generalizes our
previous results~\cite{escudero} in the following way: a fast
growth rate will make the interface weakly converge to a
hypersphere. We have further found the scaling relation
$\alpha=\beta_r z$, which quantitatively relates the center of
mass fluctuations, growth rate and rate of (weak) convergence
to/divergence from the hyperspherical profile. These results,
together with our previous ones~\cite{escudero4,escudero3},
suggest the following generalization of the critical exponents for
arbitrary values of $\gamma$
\begin{equation}
\label{gscaling} \alpha =\frac{z-d}{2}, \quad
\beta:=\frac{\alpha}{z} =\frac{1}{2}-\frac{d}{2z}, \quad
\mathrm{for} \quad z=\min \left\{z_c,\frac{1}{\gamma} \right\},
\end{equation}
which should be valid for linear models as well as some nonlinear
equations as the VLDS one, which present a conserved drift and its
critical behavior is accessible to dimensional analysis. For the
KPZ equation, which nonlinearity introduces both mass and
fluctuations into the system, we expect a different behavior. As
we have seen, the amount of mass entering the interface (and of
course the way it is introduced) may change the critical behavior.
We have explicitly shown this by means of the linear models in
absence and presence of dilution, which gave respectively rise to
correlations (\ref{radcorr1}) and (\ref{radcorr}).

\begin{table}[h]
\begin{tabular}{|c|c|c|c|}
\colrule
$\gamma < \frac{1}{z_c}$     &   $\frac{1}{z_c} < \gamma < \frac{1}{d}$   &  $\frac{1}{z_c} < \gamma < \frac{1}{d+2}$ & $\gamma > \max \left\{\frac{1}{d}, \frac{1}{z_c} \right\}$ \\
\colrule
       Classical exponents &  Roughness  &   Super-roughness & Weak convergence to the hyperspherical profile \\
\colrule
\end{tabular}
\caption{Morphologies for different spatial dimensions and
exponents. \label{table}}
\end{table}

We have summarized the different morphologies that can appear in
Table~\ref{table} (see also Fig. 2 in~\cite{escudero4}). In the
first column, for small $\gamma$, we have placed the regime in
which the classical values of the exponents are recovered. The
other three columns represent situations in which this is not the
case. The ``roughness'' column describes the situation in which
the radial fluctuations do not average each other out for long
times, so in this regime the overall appearance of the cluster
deviates from the hyperspherical profile. The ``super-roughness''
column states a particular case of this last regime, in which the
amplitude of the radial fluctuations grows faster than the
arc-length sustaining them. In both cases roughness should be
understood as the degree of weak divergence (as opposed to weak
convergence) from the hyperspherical profile. Finally, the last
column describes the regime in which the weak convergence to the
hyperspherical profile is reached asymptotically in time.

Note that the meaning of the exponent $\beta$ in~(\ref{gscaling})
changes for $\gamma < 1/z_c$ and for $\gamma > 1/z_c$. For $\gamma
< 1/z_c$ this exponent describes the short time dependence of the
spatially averaged pointwise width with time, as in the classical
situation. For $\gamma > 1/z_c$, as we have mentioned, it
describes the center of mass fluctuations, i. e., the {\it less}
dominant source of fluctuations is chosen in each regime. For
$\gamma < 1/z_c$ and $\alpha>0$ (which implies the height
difference correlation is well defined and the pointwise width is
finite) there are always center of mass fluctuations. For $\gamma
> 1/z_c$ the situation is different as it is possible to find
growth regimes with no center of mass fluctuations and which
pointwise width increases as a power law of time. This is
precisely the weak converge to the hyperspherical profile: while
microscopically the interface is rough, macroscopically it is a
concentrically propagating hypersphere. The change of meaning of
the exponent $\beta$ coincides with a change of physics. The long
range shape of the interface is no longer affected by the
diffusion mechanism, which becomes irrelevant, but for the system
size evolution, which sculptures the resulting
profile~\cite{escudero3}. And so, the violation of
scaling~(\ref{fv}) is associated with the advent of a new dominant
physical mechanism in the interface macroscale: system size growth
overtakes diffusion and determines the cluster macroscopic shape.

\section*{Acknowledgments}

This work has been partially supported by the MICINN (Spain)
through Project No. MTM2010-18128.


\begin{thebibliography} {99}
\bibitem{foot} Weak convergence (or convergence in the sense of distributions) is formalized in mathematical analysis texts as convergence in the dual
space. Intuitively, it corresponds to the spatial self-averaging
of a wildly oscillating quantity. A good example of this is given
by the Riemann-Lebesgue lemma, which states the weak convergence
of wildly oscillating trigonometric functions. Also, weak
convergence is a relevant concept in physics: it might result, for
instance, from the self-averaging of random fluctuations. As an
example, some quantum mechanical quantities weakly converge to
their classical counterparts.
\bibitem{width} The pointwise width is $W(t)= \{ \int [h(x,t)-\int h(x',t) dx']^2 dx
\}^{1/2}$. The function height $h(x,t)$ is introduced in the
following.
\end{thebibliography}
\end{document}